\documentclass{rspublic}

\usepackage{amssymb}
\usepackage{amsmath}
\usepackage{mathrsfs}
\usepackage{bm}
\usepackage{soul}

\DeclareFontShape{OT1}{cmtt}{bx}{n}{<5><6><7><8><9><10><10.95><12><14.4><17.28><20.74><24.88>cmttb10}{}

\newcommand{\beq}{\begin{equation}}
\newcommand{\eeq}{\end{equation}}

\newcommand{\tl}[1]{\widetilde{#1}}
\newcommand{\ol}[1]{\overline{#1}}

\newcommand{\p}{\partial}
\newcommand{\gtt}{\textup{\texttt{g}}}
\newcommand{\gttb}{\textup{\textbf{\texttt{g}}}}
\newcommand{\htt}{\textup{\texttt{h}}}
\newcommand{\httb}{\textup{\textbf{\texttt{h}}}}
\newcommand{\upd}{\mathrm{d}}
\newcommand{\bdag}{\boldsymbol{\dag}}
\newcounter{exple}
\addtocounter{exple}{1}
\newenvironment{example}{
\bigskip%
\small\noindent{\bf Example \arabic{exple}\quad}}{\smallskip}
\newcommand{\bex}{\begin{example}}
\newcommand{\eex}{\end{example}}

\begin{document}

\title[Noether's Second Theorem]{Extensions of Noether's Second Theorem: from continuous to discrete systems}

\author[P.E. Hydon and E.L. Mansfield]{Peter E. Hydon and Elizabeth L. Mansfield}

\affiliation{
 Dept.\ of Mathematics,
 University of Surrey, Guildford GU2 7XH, UK\\
{\it E-mail: P.Hydon@surrey.ac.uk}\\
SMSAS, University of Kent, Canterbury, Kent, CT2 7NF, UK\\
{\it E-mail: E.L.Mansfield@kent.ac.uk}
}
\label{firstpage}

\maketitle

\begin{abstract}{Conservation laws, gauge symmetries, difference equations}
A simple local proof of Noether's Second Theorem is given. This proof immediately leads to
a generalization of the theorem, yielding conservation laws and/or explicit relationships between the Euler--Lagrange equations
of any variational problem whose symmetries depend upon a set of free or partly-constrained functions.
Our approach extends further to deal with finite difference systems. The results are easy to apply; several well-known continuous and discrete systems are used as illustrations.
\end{abstract}

\section{Introduction}

Variational methods are endemic in physics and engineering. Commonly, physical symmetries are built into the Lagrangian; these correspond to conservation laws via Noether's Theorem. In classical mechanics, for instance, invariance under translation in time is associated with conservation of energy, invariance under rotations in the base space yields conservation of angular momentum, and so on. Noether's Theorem applies only to systems of Euler--Lagrange equations that are in Kovalevskaya form (see Olver (1993) for details). For other Euler--Lagrange systems, each nontrivial variational symmetry still yields a conservation law, but there is no guarantee that it will be nontrivial.

It is still not widely known that Noether's famous paper on variational problems (Noether, 1918) includes a second theorem, which applies to variational symmetries that depend on arbitrary locally smooth functions of all of the independent variables. (Henceforth, we restrict attention to functions that are locally smooth in each continuous variable.) Noether's Second Theorem states that such symmetries exist if and only if there exist differential relations between the Euler--Lagrange equations. In this case the Lie algebra of variational symmetries is infinite-dimensional and the Euler--Lagrange equations are not in Kovalevskaya form. Brading (2002) is an interesting and accessible history of the mathematics and physics of Noether's two theorems and Weyl's gauge symmetries. Brading observes that Noether's Second Theorem has received scant attention in the physics literature, despite its importance for equations that have gauge symmetries.

Noether's Theorem deals primarily with finite-dimensional Lie algebras of variational symmetry generators, whereas Noether's Second Theorem addresses the infinite-dimensional case when generators depend on arbitrary functions of \textit{all} independent variables. However, there is an intermediate case that occurs commonly: generators may depend upon functions that satisfy some constraints. For instance, the variational problem
\[
 \delta_u\int \tfrac{1}{2}\big(u_{,x}^2-u_{,t}^2\big)\, \upd x\, \upd t =0
\]
for the one-dimensional wave equation is invariant under Lie point transformations whose characteristic is of the form $\gtt(x-t)$, where $\gtt$ is an arbitrary function. Therefore the Lie algebra of variational symmetry generators is infinite-dimensional. However, the differential constraint
\[
 \gtt_{,x}+\gtt_{,t}=0
\]
applies; there are no generators that depend upon an unconstrained arbitrary function $\gtt(x,t)$. Rosenhaus (2002) examines this type of problem from the global viewpoint, determining which spatial boundary conditions are necessary for the existence of integral quantities that are conserved in time. However, the effect of constraints on local conservation laws has remained an open problem until now.

Some important variational problems have variational symmetry generators that depend on several arbitrary functions which are linked by constraints. Noether's Second Theorem has not yet been generalized to deal with this type of problem.

A discrete variational problem may be a model of a discrete process, or it may be a discretization of a continuous variational problem. Either way, discrete analogues
of Noether's two theorems are important. Noether's Theorem for finite difference variational problems has now been well-studied; see, for example, Dorodnitsyn (2001), Hickman \& Hereman (2003) and
Hydon \& Mansfield (2004). If a discretization preserves a variational symmetry, the discrete version of
Noether's Theorem guarantees that the corresponding conservation law is also preserved.

Christiansen \& Halvorsen (2011) have constructed a gauge symmetry-preserving discretization of the Maxwell--Klein--Gordon equations for a charged particle in an electromagnetic field. They used the discrete analogue of Noether's Theorem to obtain a linear relationship between the discrete Euler--Lagrange equations; this was viewed (correctly) as the satisfaction of a physically-important constraint. So there is evidently a need for a discrete analogue of Noether's Second Theorem.

The current paper has three purposes. First, we give a simple local proof of Noether's Second Theorem for a single arbitrary function and extend this to deal with variational symmetries that depend on several arbitrary functions. Then we show what happens locally when the arbitrary functions are constrained in some way. Finally, we transfer these ideas to discrete variational problems. Some well-known systems are used to illustrate the application of our results.

\section{Symmetries of Lagrangians: the continuous case}

We begin by summarizing a few key facts about variational symmetries; for details, see Olver (1993).
We work locally, using independent variables $\mathbf{x} = (x^1,\dots, x^p)$ and dependent variables $\mathbf{u}=(u^1,\dots,u^q)$. Derivatives of each $u^\alpha$ are written in the form $u^\alpha_{,\mathbf{J}}$, where $\mathbf{J}=(j^1,\dots,j^p)$ is a multi-index; here each $j^i$ is a non-negative integer that denotes the number of derivatives with respect to $x^i$, and so $u^\alpha_{,\mathbf{0}}\equiv u^\alpha$. The \textit{total derivative} with respect to $x^i$ is the operator
\[
D_i=\frac{\p}{\p x^i}+\sum_{\alpha,\mathbf{J}}\frac{\p u^\alpha_{,\mathbf{J}}}{\p x^i}\frac{\p}{\p u^\alpha_{,\mathbf{J}}}\,,
\]
and we shall use the following shorthand:
\[
\mathbf{D}_\mathbf{J}=D_1^{j^1}D_2^{j^2}\dots D_p^{j^p},\qquad (\mathbf{D}_\mathbf{J})^{\bdag}=(-\mathbf{D})_\mathbf{J}=(-1)^{j^1+\cdots+j^p}\mathbf{D}_\mathbf{J}.
\]
Here $\bdag$ denotes the formal adjoint.
Henceforth, we adopt the Einstein summation convention to avoid a plethora of summation symbols.
Consider the action
\[
\mathcal{L}[\mathbf{u}] = \int L(\mathbf{x},[\mathbf{u}])\,\upd \mathbf{x},
\]
where $[\mathbf{u}]$ in the integrand is shorthand for $\mathbf{u}$ and finitely many of its derivatives. The Euler--Lagrange equations are obtained by varying each $u^\alpha$ and integrating by parts:
\beq\label{oldfunc}
0=\delta_{\mathbf{u}}\mathcal {L}[\mathbf{u}]\equiv\int\frac{\p L}{\p u^\alpha_{,\mathbf{J}}}\delta u^\alpha_{,\mathbf{J}}\,\upd \mathbf{x}=\int\frac{\p L}{\p u^\alpha_{,\mathbf{J}}}\mathbf{D}_\mathbf{J}(\delta u^\alpha)\,\upd \mathbf{x}=\int \mathbf{E}_\alpha (L) \delta u^\alpha \,\upd \mathbf{x}.
\eeq
Here the Euler operator $\mathbf{E}_\alpha$, which corresponds to variations in $u^\alpha$, is defined by
\[
\mathbf{E}_\alpha (L)=(-\mathbf{D})_\mathbf{J}\left(\frac{\p L}{\p u^\alpha_{,\mathbf{J}}}\right).
\]
Following Olver, we have assumed that variations are restricted to ensure that there are no contributions from the boundary of the domain. With this formal approach, two Lagrangians $L$ and $L'$ lead to the same set of Euler--Lagrange equations,
\beq\label{ELbasic}
\mathbf{E}_\alpha(L) =0,\qquad \alpha = 1,\dots,q,
\eeq
if and only if $L-L'$ is a total divergence.

Typically, the set of all generalized (or Lie--B\"{a}cklund) symmetries of the Euler--Lagrange equations (\ref{ELbasic}) is not a group. However, just as for point symmetries, one can find infinitesimal generators (in evolutionary form, prolonged to all orders),
\[
 X=Q^\alpha\frac{\p }{\p u^\alpha}+(D_iQ^\alpha)\frac{\p }{\p D_i u^\alpha}+\dots= \big(\mathbf{D}_{\mathbf{J}}Q^{\alpha}\big)\frac{\p}{\p u^\alpha_{,\mathbf{J}}},
\]
by solving the linearized symmetry condition,
\[
X\big(\mathbf{E}_\beta(L)\big)=0\quad\text{on solutions of (\ref{ELbasic})},\qquad \beta=1,\dots,q.
\]
The solution is a $q$-tuple $\mathbf{Q(\mathbf{x},[\mathbf{u}])}=(Q^1,\dots,Q^q)$, which is called the characteristic of the symmetries generated by $X$. The set of characteristics of all generalized symmetries of the Euler--Lagrange equations (\ref{ELbasic}) is a Lie algebra, whose Lie bracket is
\beq\label{Lieprod}
[\mathbf{Q}_1,\mathbf{Q}_2]=X_1(\mathbf{Q}_2)-X_2(\mathbf{Q}_1).
\eeq

Variational symmetries (in the broadest sense) are generalized symmetries whose infinitesimal generators satisfy the additional condition
\beq\label{xldiv}
 X(L) = D_i\Big(P_0^i(\mathbf{x},[\mathbf{u}])\Big),
\eeq
for some functions $P_0^1,\dots, P_0^p$. This condition ensures that the variational problem is invariant under such symmetries. The set of characteristics of all variational symmetries is a Lie algebra with the Lie bracket (\ref{Lieprod}). Integrating (\ref{xldiv}) by parts yields an expression of the form
\beq
 Q^{\alpha}\mathbf{E}_{\alpha}(L) = D_i\big(P^i(\mathbf{x},[\mathbf{u}])\big).
\label{first}
\eeq
Consequently, each variational symmetry characteristic $\mathbf{Q}$ yields a conservation law
\beq\label{N1CL}
D_i\big(P^i(\mathbf{x},[\mathbf{u}])\big)=0\quad\text{on solutions of (\ref{ELbasic})}.
\eeq
This is the local version of Noether's first (and best-known) theorem on variational symmetries.

\section{A local approach to Noether's Second Theorem}

Olver (1993) states and proves Noether's Second Theorem in the following form (using our notation).

\begin{theorem}  The variational problem (\ref{oldfunc}) admits an infinite-dimensional group of variational symmetries whose characteristics $\mathbf{Q}(\mathbf{x},[\mathbf{u};\gtt])$ depend on an arbitrary function $\gtt(\mathbf{x})$ (and its derivatives) if and only if there exist differential operators $\mathcal{D}^1,\dots,\mathcal{D}^q$, not all zero, such that
\beq\label{N2Ol}
 \mathcal{D}^1\mathbf{E}_1(L) + \cdots + \mathcal{D}^q\mathbf{E}_q(L)\equiv 0
\eeq
for all $\mathbf{x}, \mathbf{u}$.
\end{theorem}

The theorem as stated is restricted to variational problems whose variational symmetries \textit{do} form a group; such problems are uncommon. However, if `group of variational symmetries' is replaced by `Lie algebra of variational symmetry generators,' the theorem applies to all variational problems (\ref{oldfunc}). We shall use the more general wording henceforth.

The `if' part of Olver's proof is straightforward. Multiply (\ref{N2Ol}) by an arbitrary function $\gtt(\mathbf{x})$ and integrate by parts to get an expression of the form (\ref{first}), where
\[
Q^\alpha(\mathbf{x},[\mathbf{u};\gtt])=\big(\mathcal{D}^\alpha\big)^{\bdag} (\gtt).
\]
Then reverse the steps that lead from (\ref{xldiv}) to (\ref{first}) to show that $\mathbf{Q}$ is the characteristic of a variational symmetry for each $\gtt(\mathbf{x})$. Consequently, if there are no characteristics of variational symmetries that depend on a completely arbitrary function of $\mathbf{x}$, no relation of the form (\ref{N2Ol}) exists.

The `only if' part of the proof is quite long, so we do not reproduce it here. Instead, we use the following constructive proof, which is much simpler. The key is to regard $\gtt$ as an additional dependent variable and to apply the Euler--Lagrange operator $\mathbf{E}_\gtt$ to (\ref{first}), giving
\beq\label{EgQE}
\mathbf{E}_\gtt\big\{Q^\alpha(\mathbf{x},[\mathbf{u};\gtt])\mathbf{E}_{\alpha}(L)\big\}\equiv 0.
\eeq
This is the required differential relation between the Euler--Lagrange equations. Explicitly, it amounts to
\beq\label{diffrel1}
(-\mathbf{D})_\mathbf{J}\left(\frac{\p Q^\alpha(\mathbf{x},[\mathbf{u};\gtt])}{\p \gtt_{,\mathbf{J}}}\,\mathbf{E}_{\alpha}(L)\right)\equiv 0.
\eeq
If $\mathbf{Q}$ is linear in $\gtt$ and its derivatives, this relation is independent of $\gtt$. Otherwise, one can linearize $\mathbf{Q}$ by taking its Fr\'{e}chet derivative with respect to $\gtt$, as explained by Olver; we discuss an alternative approach later in this section.
Typically the characteristic is independent of high-order derivatives of $\gtt$ and so (\ref{diffrel1}) is easy to determine; the following example illustrates this.

[Note: in each example we use whatever notation is standard, so that readers can see the results in a familiar form, without having to translate from the notation that we use for the general theory.]

\bex\refstepcounter{exple}\label{gaugeex}
The interaction of a scalar particle of mass $m$ and charge $e$ with an electromagnetic field has a variational formulation with well-known gauge symmetries.
The independent variables are the standard flat space-time coordinates $\mathbf{x}=(x^0, x^1, x^2, x^3)$, where $x^0$ denotes time. The dependent variables are the complex-valued scalar $\psi$ (which is the wavefunction for the particle), its complex conjugate, $\psi^*$,
and the real-valued electromagnetic four-potential $A^{\mu}$, $\mu=0,1,2,3$; indices are raised and lowered using the metric $\eta=\text{diag}\{-1,1,1,1\}$. The Lagrangian is
\begin{equation}\label{emLag}
L= \frac14 F_{\mu\nu}F^{\mu\nu} + \left(\nabla_{\mu}\psi\right)\left(\nabla^{\mu}\psi\right)^*+m^2\psi\psi^*
\end{equation}
where
$$F_{\mu\nu}=A_{\nu,\mu}-A_{\mu,\nu},\qquad \nabla_{\mu}=D_{\mu}+\ri e A_{\mu}.$$
Therefore the Euler--Lagrange equations are
\begin{align*}
 0&= \mathbf{E}_\psi (L) \equiv -\left(\nabla_\mu\nabla^\mu\psi\right)^{\! *}+m^2\psi^*,\\
 0&= \mathbf{E}_{\psi^*}(L) \equiv -\nabla_\mu\nabla^\mu\psi+m^2\psi,\\
 0&= \mathbf{E}_\sigma (L) \equiv \ \ri e\psi\left(\nabla_\sigma \psi\right)^{\! *} - \ri e\psi^*\nabla_\sigma\psi+\eta_{\sigma\alpha}F^{\alpha\beta}_{,\beta},
\end{align*}
where $\mathbf{E}_\sigma$ is obtained by varying $A^\sigma$.
The variational symmetries include the gauge symmetries
$$\psi\mapsto \exp(-\ri e\lambda)\psi,\qquad \psi^*\mapsto \exp(\ri e\lambda)\psi^*,\qquad A^{\sigma}\mapsto A^{\sigma}+\eta^{\sigma\alpha}\lambda_{,\alpha},$$
where $\lambda$ is an arbitrary real-valued function of $\mathbf{x}$; the Lagrangian $L$ is invariant under these transformations, so $XL=0$.
The characteristics of the gauge symmetries have components
$$Q^{\psi}=-\ri e \psi \gtt,\qquad Q^{\psi^*}=\ri e \psi^* \gtt,\qquad Q^{\sigma}=\eta^{\sigma\alpha}\gtt_{,\alpha},$$
where $\gtt$ is an arbitrary real-valued function of $\mathbf{x}$. Consequently
\[
 \mathbf{E}_\gtt\big\{ Q^{\psi}\mathbf{E}_\psi (L)+ Q^{\psi^*}\mathbf{E}_{\psi^*}(L) +Q^{\sigma} \mathbf{E}_\sigma (L)\big\}=-\ri e\psi\mathbf{E}_\psi (L)+\ri e\psi^*\mathbf{E}_{\psi^*}(L)-D_\alpha\left(\eta^{\sigma\alpha} \mathbf{E}_\sigma (L)\right),
\]
and so Noether's Second Theorem yields the differential relation
\[
-\ri e\psi\mathbf{E}_\psi (L)+\ri e\psi^*\mathbf{E}_{\psi^*}(L)-D_\alpha\left(\eta^{\sigma\alpha} \mathbf{E}_\sigma (L)\right)\equiv 0.
\]
\eex

Our proof of the `only if' part of Noether's Second Theorem is essentially a local version of Noether's original proof (see Noether, 1918). Like Noether's proof, it extends immediately to variational symmetries whose characteristics depend on $R$ independent arbitrary functions $\gttb=(\gtt^1(\mathbf{x}),\dots, \gtt^R(\mathbf{x}))$ and their derivatives. This gives $R$ differential relations between the Euler--Lagrange equations:
\beq\label{EgrQE}
\mathbf{E}_{\gtt^r}\big\{Q^\alpha(\mathbf{x},[\mathbf{u};\gttb])\mathbf{E}_{\alpha}(L)\big\}\equiv(-\mathbf{D})_\mathbf{J}\left(\frac{\p Q^\alpha(\mathbf{x},[\mathbf{u};\gttb])}{\p \gtt^r_{,\mathbf{J}}}\,\mathbf{E}_{\alpha}(L)\right)= 0,\quad r=1,\dots,R.
\eeq

A useful observation is that this set of relations is invariant under any locally invertible change of the arbitrary functions. For suppose that each $\gttb^r$ is a locally invertible function of a new set of arbitrary functions $\htt^\rho,\ \rho=1,\dots,R$. Then the identity
\beq\label{EhrQE}
\mathbf{E}_{\htt^\rho}\big\{Q^\alpha(\mathbf{x},[\mathbf{u};\gttb(\httb)])\mathbf{E}_{\alpha}(L)\big\}\equiv
\frac{\p \gtt^r(\httb)}{\p \htt^\rho}\Big(\mathbf{E}_{\gtt^r}\big\{Q^\alpha(\mathbf{x},[\mathbf{u};\gttb])\mathbf{E}_{\alpha}(L)\big\}\Big) \Big{|}_{\gttb=\gttb(\httb)}
\eeq
shows that the set of relations
\[
\mathbf{E}_{\htt^\rho}\big\{Q^\alpha(\mathbf{x},[\mathbf{u};\gttb(\httb)])\mathbf{E}_{\alpha}(L)\big\}=0, \qquad \rho=1,\dots,R,
\]
is obtained by taking an invertible linear combination of the original relations (\ref{EgrQE}). So if there exists a locally invertible change of variables such that $\mathbf{Q}$ is linear in $\httb$, (\ref{EgrQE}) is equivalent to a set of relations that is independent of any arbitrary functions.

It is convenient to regard (\ref{EgrQE}) as `new' Euler--Lagrange equations that arise when $\mathbf{g}$ is varied in the functional
\beq\label{newfunc}
\hat{\mathcal{L}}[\mathbf{u};\mathbf{g}] = \int \hat{L}(\mathbf{x},[\mathbf{u};\mathbf{g}])\,\upd \mathbf{x},
\eeq
where
\beq\label{newL}
\hat{L}(\mathbf{x},[\mathbf{u};\mathbf{g}])=Q^\alpha(\mathbf{x},[\mathbf{u};\mathbf{g}])\mathbf{E}_{\alpha}(L(\mathbf{x},[\mathbf{u}])).
\eeq
To prove that the relations (\ref{EgrQE}) are independent, suppose that the converse is true; then there exist differential operators $\hat{\mathcal{D}}^r$ such that
\[
 \hat{\mathcal{D}}^1\mathbf{E}_{\gtt^1}(\hat{L}) + \cdots + \hat{\mathcal{D}}^R\mathbf{E}_{\gtt^R}(\hat{L})\equiv 0.
\]
By the `if' part of Noether's Second Theorem, there exist variational symmetries of (\ref{newfunc}) whose generator is of the form
\[
\hat{X}=D_{\mathbf{J}}\Big\{\big(\hat{\mathcal{D}}^r\big)^{\bdag} (\hat{\gtt})\Big\}\frac{\p}{\p \gtt^r_{,\mathbf{J}}}\,,
\]
where $\hat{\gtt}$ is an arbitrary function. In other words, there exist functions $\hat{P}^i(\mathbf{x},[\mathbf{u};\gttb;\hat{\gtt}])$ such that
\[
\hat{X}\Big\{ Q^\alpha(\mathbf{x},[\mathbf{u};\gttb])\Big\}\mathbf{E}_{\alpha}(L(\mathbf{x},[\mathbf{u}]))=\hat{X}\big(\hat{L}\big)=D_i\big(\hat{P}^i\big).
\]
Consequently, for arbitrary $\gttb$ and $\hat{\gtt}$,
\[
\hat{Q}^\alpha(\mathbf{x},[\mathbf{u};\gttb;\hat{\gtt}])=\hat{X}\Big\{ Q^\alpha(\mathbf{x},[\mathbf{u};\gttb])\Big\}
\]
is a characteristic of variational symmetries for the original variational problem (\ref{oldfunc}). This implies that the set of variational symmetries for (\ref{oldfunc}) depends on more than $R$ independent arbitrary functions, contradicting our original assumption. Hence the relations (\ref{EgrQE}) cannot be dependent.

Conversely, given $R$ independent relations between the Euler--Lagrange equations, Olver's `if' proof applied to each one shows that there are characteristics of variational symmetries that depend on $R$ independent arbitrary functions, namely
\beq\label{grchar}
Q^\alpha(\mathbf{x},[\mathbf{u};\gttb])=\big(\mathcal{D}^\alpha_r\big)^{\bdag} (\gtt^r).
\eeq
Thus we arrive at the general form of Noether's Second Theorem.

\begin{theorem}  The variational problem (\ref{oldfunc}) admits an infinite-dimensional Lie algebra of variational symmetry generators whose characteristics $\mathbf{Q}(\mathbf{x},[\mathbf{u};\gttb])$ depend on $R$ independent arbitrary functions $\gttb=(\gtt^1(\mathbf{x}),\dots, \gtt^R(\mathbf{x}))$ and their derivatives if and only if there exist differential operators $\mathcal{D}^\alpha_r$ that yield $R$ independent differential relations,
\beq\label{N2HM1}
 \mathcal{D}^\alpha_r\mathbf{E}_\alpha(L)\equiv 0,\qquad r=1,\dots,R,
\eeq
between the Euler--Lagrange equations. Given the relations (\ref{N2HM1}), the corresponding characteristics are (\ref{grchar})
Conversely, given the characteristics, the corresponding differential relations are (\ref{EgrQE}).
\end{theorem}

\section{Constrained variational symmetries}
Constraints on arbitrary functions in $\mathbf{Q}$ arise from the linearized symmetry condition for the Euler--Lagrange equation, coupled with the (linear) requirement that the symmetries are variational. Therefore we now suppose that the characteristics of variational symmetries of (\ref{oldfunc}) depend on functions $\gttb$ that are subject to $S$ linear differential constraints,
\beq\label{constr}
\mathscr{D}_{sr}(\gtt^r)=0,\qquad s=1,\dots,S,
\eeq
where each $\mathscr{D}_{sr}$ is a differential operator. We assume that this set of constraints is complete, in the sense that it yields no additional integrability conditions.
The constraints can be incorporated into the Lagrangian (\ref{newL}), yielding
\beq\label{newLcon}
\hat{L}(\mathbf{x},[\mathbf{u};\gttb])=Q^\alpha(\mathbf{x},[\mathbf{u};\gttb])\mathbf{E}_{\alpha}(L(\mathbf{x},[\mathbf{u}]))-
\nu^s\mathscr{D}_{sr}(\gtt^r),
\eeq
where $\nu^1,\dots,\nu^s$ are Lagrange multipliers. Taking variations of (\ref{newfunc}) with respect to $\gttb$, we obtain
\beq\label{EgQEcon}
(-\mathbf{D})_\mathbf{J}\left(\frac{\p Q^\alpha(\mathbf{x},[\mathbf{u};\gttb])}{\p \gtt^r_{,\mathbf{J}}}\,\mathbf{E}_{\alpha}(L)\right)\equiv(\mathscr{D}_{sr})^{\bdag}(\nu^s),\qquad r=1,\dots,R.
\eeq

If $S<R$, one may be able to eliminate the undetermined functions $\nu^s$ from (\ref{EgQEcon}), as will be illustrated by the next example. The elimination of a particular set ${\cal S}$ of dependent variables from a differential system may be accomplished using algorithms in differential algebra. Hubert (2000) and Hubert (2003) provide a careful and extensive summary of the state of the art to date. Differential elimination algorithms have been implemented in the {\sc Maple} library packages {\tt diffalg} and
{\tt rifsimp}. These algorithms come with a `certificate'; provided that the correct elimination term ordering is
used, the elimination ideal is generated by those equations ${\cal E}$ in the output that are independent of the variables in ${\cal S}$. This means that any equation that is algebraically or differentially derived from the
input set, and is free of the elements of ${\cal S}$, can be derived from ${\cal E}$. If ${\cal E}$ is empty then the elimination ideal
itself is empty; in this case, none of the equations that can be derived from the input set are independent of all variables in ${\cal S}$.

Incidentally, the elimination algorithm can be used to test directly whether or not an Euler--Lagrange system
has a differential relation such as (\ref{N2HM1}): simply replace $\{E^{\alpha}(L)=0\}$
with $\{E^{\alpha}(L)-f^{\alpha}=0\}$ and eliminate  ${\cal S}=\{u^{\alpha}\}$. The resulting equations for $\{f^{\alpha}\}$ codify the ideal of differential relations on the expressions $\{E^{\alpha}(L)\}$.

There are subtleties for nonlinear systems; indeed, the proof that the output of the algorithms has
the stated properties is one of the most subtle and complex problems of differential algebra.

\bex\refstepcounter{exple}\label{twoarbs}
Suppose that $\mathbf{Q}$ depends on two functions $\gtt^1(x^1,x^2)$ and $\gtt^2(x^1,x^2)$ that are subject to the single constraint
\beq
\gtt^1_{,1} + \gtt^2_{,2} =0
\label{tenth}
\eeq
(This type of constraint occurs when the variational symmetries include all area-preserving diffeomorphisms of the plane.)
Then (\ref{EgQEcon}) amounts to
\begin{eqnarray*}
(-\mathbf{D})_\mathbf{J}\left(\frac{\p Q^\alpha(\mathbf{x},[\mathbf{u};\gttb])}{\p \gtt^1_{,\mathbf{J}}}\,\mathbf{E}_{\alpha}(L)\right)&\equiv& -\nu_{,1}, \\
(-\mathbf{D})_\mathbf{J}\left(\frac{\p Q^\alpha(\mathbf{x},[\mathbf{u};\gttb])}{\p \gtt^2_{,\mathbf{J}}}\,\mathbf{E}_{\alpha}(L)\right)&\equiv& -\nu_{,2}.
\end{eqnarray*}
By eliminating the Lagrange multiplier $\nu$, we obtain a single differential relation between the Euler--Lagrange equations:
\beq
(-D_2)(-\mathbf{D})_\mathbf{J}\left(\frac{\p Q^\alpha(\mathbf{x},[\mathbf{u};\gttb])}{\p \gtt^1_{,\mathbf{J}}}\,\mathbf{E}_{\alpha}(L)\right) + D_1 (-\mathbf{D})_\mathbf{J}\left(\frac{\p Q^\alpha(\mathbf{x},[\mathbf{u};\gttb])}{\p \gtt^2_{,\mathbf{J}}}\,\mathbf{E}_{\alpha}(L)\right)\equiv 0.
\label{eleventh}
\eeq
This is precisely the same constraint as would be obtained by using the local solution of (\ref{tenth}),
\[
 \gtt^1=\gtt_{,2},\quad \gtt^2 =-\gtt_{,1}
\]
(where $\gtt$ is entirely arbitrary) in Noether's Second Theorem.  However the Lagrange multiplier method has the advantage of working even when (\ref{constr}) cannot be solved explicitly.\eex

When the system of constraints (\ref{constr}) does not allow the multipliers $\nu^s$ to be eliminated, the most that one can do is to construct conservation laws of the Euler--Lagrange equations, as follows. First determine a particular solution of (\ref{EgQEcon}) for $\nu^s$; this can be done by inspection (which is fast), or by the homotopy method (which is systematic). The resulting solution is then substituted into
\beq\label{gclaw}
\nu^s\mathscr{D}_{sr}(\gtt^r)-\gtt^r(\mathscr{D}_{sr})^{\bdag}(\nu^s)=0\qquad\text{when}\quad \mathbf{E}_{1}(L)=\cdots=\mathbf{E}_{q}(L)=0,
\eeq
yielding a set of conservation laws that depends on $\gttb$. This set is independent of which particular solution is used. It can be shown that if $\mathbf{Q}$ is homogeneous with respect to $\gttb$, the conservation law (\ref{gclaw}) is a multiple of the one that is obtained from Noether's (first) Theorem.

\bex\refstepcounter{exple}\label{weex}
As we discussed in the Introduction, the Lagrangian
\[
L=\tfrac{1}{2}\big((u_{,x})^2-(u_{,t})^2\big)
\]
for the one-dimensional scalar wave equation,
\beq\label{wave}
u_{,tt}-u_{,xx}=0,
\eeq
admits variational symmetries with partly-constrained characteristics. Indeed, the most general variational symmetry characteristic is $Q=\gtt^1(x,t)+\gtt^2(x,t)$, where
\[
 \gtt^1_{,x}+\gtt^1_{,t}=0,\qquad \gtt^2_{,x}-\gtt^2_{,t}=0.
\]
So the modified Lagrangian (\ref{newLcon}) is
\[
\hat{L}=(u_{,tt}-u_{,xx})(\gtt^1+\gtt^2)-\nu^1(\gtt^1_{,x}+\gtt^1_{,t})-\nu^2(\gtt^2_{,x}-\gtt^2_{,t}).
\]
\eex
Taking variations with respect to $\gttb=(\gtt^1,\gtt^2)$ gives the identities
\[
u_{,tt}-u_{,xx}\equiv-(\nu^1_{,x}+\nu^1_{,t}),\qquad u_{,tt}-u_{,xx}\equiv-(\nu^2_{,x}-\nu^2_{,t}),
\]
which are satisfied by
\[
\nu^1=u_{,x}-u_{,t},\qquad \nu^2=u_{,x}+u_{,t}.
\]
Then the conservation laws (\ref{gclaw}) amount to
\[
\big\{(\gtt^1+\gtt^2)u_{,x}-(\gtt^1-\gtt^2)u_{,t}\big\}_{,x}+\big\{(\gtt^1-\gtt^2)u_{,x}-(\gtt^1+\gtt^2)u_{,t}\big\}_{,t}=0\ \, \text{when (\ref{wave}) holds}.
\]
There are no trivial conservation laws with $Q$ nonzero.

\bex\refstepcounter{exple}\label{swwex} For a more substantial example that produces interesting conservation laws, consider the two-dimensional shallow water equations from Lagrangian fluid mechanics.
Each fluid particle is labelled by its position $(a^1,a^2)$ at some reference time, $t=0$, and $(x,y)=(x(a^1,a^2,t),y(a^1,a^2,t))$ is the position of
the fluid particle at time $t$. We set $\dot{x}=x_{,t} $ and $\dot{y}=y_{,t}$; from here on only the derivatives with respect to the label variables $a^i$ are denoted by subscripts.
Salmon (1983) introduced a Lagrangian for the shallow water equations; with constant Coriolis parameter $f$, it amounts to 
$$ L([x,y])= \frac12 \left (\dot{x}^2+\dot{y}^2\right) + fx\dot{y}-\frac12 g h,$$
where $$ h=\frac1{x_{,1}y_{,2}-x_{,2}y_{,1}},$$ is the fluid depth and
$g$ is the constant of gravity. To translate between the Eulerian and Lagrangian viewpoints, we need the identities
$$\frac{\partial}{\partial x}= h\left( y_{,2}\frac{\partial}{\partial a^1}-y_{,1}\frac{\partial}{\partial a^2}\right),\qquad
\frac{\partial}{\partial y}= h\left( -x_{,2}\frac{\partial}{\partial a^1}+x_{,1}\frac{\partial}{\partial a^2}\right).$$
Calculating the Euler--Lagrange system yields the shallow water equations
\begin{equation}\label{sww}
E_x(L)\equiv-\ddot{x} + f \dot{y} - g \frac{\partial h}{\partial x}=0,\qquad E_y(L)\equiv-\ddot{y} - f \dot{x} - g \frac{\partial h}{\partial y}=0.\end{equation}
These are supplemented by the continuity equation,
\[
\dot{h}+ h\left(\frac{\partial \dot{x}}{\partial x}+\frac{\partial \dot{y}}{\partial y}\right)=0,
\]
which is {\em not} an Euler--Lagrange equation in Salmon's formulation.
The variational symmetries include the well-known particle relabelling symmetries; these are arbitrary area-preserving diffeomorphisms of label space, whose characteristics are (Bila, Mansfield \& Clarkson, 2006)
$$ Q^x = x_{,i}\gtt^i,\qquad Q^y = y_{,i}\gtt^i,\qquad \text{where}\quad\dot{\gtt}^1=\dot{\gtt}^2=\gtt^1_{,1}+\gtt^2_{,2}=0.$$
Taking the three constraints on the functions $\gttb=(\gtt^1,\gtt^2)$ into account, the modified Lagrangian is
$$\hat{L} =  x_{,i}\gtt^iE_x(L) + y_{,i}\gtt^iE_y(L) -\nu^1\dot{\gtt}^1 -\nu^2\dot{\gtt}^2 -\nu^3\left(\gtt^1_{,1}+\gtt^2_{,2}\right).$$
Taking variations with respect to $\gttb$, we obtain
$$x_{,i} E_x(L)+y_{,i} E_y(L) \equiv -\dot{\nu}^i-\nu^3_{,i}, \qquad i=1,2,$$
which is satisfied by
\begin{align*}
 \nu^1&=\dot{x}x_{,1}+\dot{y}y_{,1}+fxy_{,1},\\
 \nu^2&=\dot{x}x_{,2}+\dot{y}y_{,2}+fxy_{,2},\\
 \nu^3&=-\tfrac{1}{2} (\dot{x}^2+\dot{y}^2) -fx\dot{y} +gh.
\end{align*}
With these solutions $\nu^i$, the resulting conservation laws (\ref{gclaw}) are
\[
 D_t(\gtt^1\nu^1+\gtt^2\nu^2)+(\gtt^1\nu^3)_{,1}+(\gtt^2\nu^3)_{,2}=0\qquad \text{when (\ref{sww}) holds}.
\]
In particular, taking $(\gtt^1,\gtt^2)$ to be $(1,0)$ and $(0,1)$ in turn yields
\[
 \dot{\nu}^1+\nu^3_{,1}=0,\qquad \dot{\nu}^2+\nu^3_{,2}=0\qquad \text{when (\ref{sww}) holds},
\]
which were discovered in a (somewhat complicated) multisymplectic setting by Hydon (2005). Differentiating the first of these with respect to $a^2$, the second with respect to $a^1$ and eliminating $\nu^3_{,12}$, we obtain the well-known conservation law for potential vorticity,
\[
 D_t\left\{\frac{1}{h}\left(\frac{\partial \dot{y}}{\partial x}-\frac{\partial \dot{x}}{\partial y} +f\right)\right\}=0\qquad \text{when (\ref{sww}) holds}.
\]
So in this example, our extension of Noether's Second Theorem has led quickly and easily to some fundamental conservation laws. Conservation of potential vorticity also arises directly from Noether's Theorem, but considerable effort is needed to obtain it (see Bila, Mansfield \& Clarkson, 2006).
\eex

\section{From continuous to discrete: Noether's Second Theorem}\label{sec:dis}

For difference equations, the dependent variables are again $\mathbf{u} = (u^1,\dots,u^q)$, but now the independent variables are $\mathbf{n} = (n^1,\dots,n^p)$, where each $n^i$ is an integer. The forward shift operators, $S_i$, are defined by
\[
 S_i:n^j\mapsto n^j+\delta_i^j,\qquad S_i f(\mathbf{n})=f(S_i\mathbf{n}),
\]
for all functions $f$; here $\delta_i^j$ is the Kronecker delta. The adjoint of $S_i$ is the backward shift $(S_i)^{\bdag}=S_i^{-1}:n^j\mapsto n^j-\delta_i^j$. We also use the multiple shifts, $\mathbf{S}_{\mathbf{J}} = S_1^{j_1}\cdots S_p^{j_p}$, and the forward difference operators,
\[
\tl{D}_i=S_i-\text{id},
\]
where $\text{id}$ denotes the identity map. (Here and in what follows, we use notation that is similar to the continuous case in order to highlight the similarities. To avoid confusion, we use a tilde to distinguish an operator from its continuous counterpart.)

For difference variational problems, the action is of the form
\[
\mathcal{L}[\mathbf{u}] = \sum_{\mathbf{n}}L\big(\mathbf{n},[\mathbf{u}]\big),
\]
where now $[\mathbf{u}]$ denotes $\mathbf{u}(\mathbf{n})$ and finitely many of its shifts. Then the Euler--Lagrange equations are
\beq
 \tl{\mathbf{E}}_{\alpha}(L)\equiv \mathbf{S}_{-\mathbf{J}}\left\{\frac{\p L}{\p \mathbf{S}_{\mathbf{J}}u^{\alpha}}\right\}=0,\qquad \alpha =1,\dots,q.
\label{twelve}
\eeq
Generalized symmetries of the Euler--Lagrange equations have generators of the form
$$\tl{X}=\mathbf{S}_{\mathbf{J}}\big(Q^{\alpha}(\mathbf{n},[\mathbf{u}])\big) \frac{\partial}{\partial \mathbf{S}_{\mathbf{J}}u^{\alpha}}\,$$
that satisfy the linearized symmetry condition,
\[
\tl{X}\big(\tl{\mathbf{E}}_{\alpha}(L)\big) =0\qquad\text{when (\ref{twelve}) holds}.
\]
Here $Q^{\alpha}(\mathbf{n},[\mathbf{u}]),\ \alpha = 1,\dots,q$, are the components of the characteristic, $\mathbf{Q}(\mathbf{n},[\mathbf{u}])$.
The symmetries are \textit{variational} if they do not change the action, that is, if there exist functions $P_0^i\big(\mathbf{n},[\mathbf{u}]\big)$ such that
\beq
\tl{X}(L) \equiv \tl{D}_i P_0^i\big(\mathbf{n},[\mathbf{u}]\big),
\label{fourteen}
\eeq
Summing (\ref{fourteen}) by parts yields the following condition, which is analogous to (\ref{first}):
\beq
Q^{\alpha}\tl{\mathbf{E}}_{\alpha}(L) = \tl{D}_i  P^i\big(\mathbf{n},[\mathbf{u}]\big),
\label{fifteen}
\eeq
for some functions $ P^i$. So Noether's Theorem applies just as in the continuous case: every variational symmetry characteristic yields a conservation law of the Euler--Lagrange equations, namely
\[
\tl{D}_i  P^i\big(\mathbf{n},[\mathbf{u}]\big)=0,\qquad\text{when (\ref{twelve}) holds}.
\]

With the above notation, the analogue of Noether's Second Theorem for the difference calculus of
variations is as follows.

\begin{theorem}  The difference variational problem $\delta_\mathbf{u}\tl{\mathcal{L}}[\mathbf{u}] =0$ admits an infinite-dimensional Lie algebra of variational symmetry generators whose characteristics $\tl{\mathbf{Q}}(\mathbf{n},[\mathbf{u};\gttb])$ depend on $R$ independent arbitrary functions $\gttb=(\gtt^1(\mathbf{n}),\dots, \gtt^R(\mathbf{n}))$ and their derivatives if and only if there exist difference operators $\tl{\mathcal{D}}^\alpha_r$ that yield $R$ independent difference relations,
\beq\label{N2HM2}
 \tl{\mathcal{D}}^\alpha_r\tl{\mathbf{E}}_\alpha(L)\equiv 0,\qquad r=1,\dots,R,
\eeq
between the Euler--Lagrange equations. Given the relations (\ref{N2HM2}), the corresponding characteristics are
\beq
Q^\alpha(\mathbf{n},[\mathbf{u};\gttb])=\big(\tl{\mathcal{D}}^\alpha_r\big)^{\bdag} (\gtt^r).
\eeq
Conversely, given the characteristics, the corresponding difference relations are
\beq\label{EgrQEd}
 \tl{\mathbf{E}}_{\gtt^r}\big\{Q^\alpha(\mathbf{x},[\mathbf{u};\gttb])\tl{\mathbf{E}}_{\alpha}(L)\big\}\equiv \mathbf{S}_{-\mathbf{J}}\left(\frac{\p Q^\alpha(\mathbf{x},[\mathbf{u};\gttb])}{\p \mathbf{S}_\mathbf{J}\gtt^r}\,\tl{\mathbf{E}}_{\alpha}(L)\right)= 0,
\eeq
and consequently
\[
\tl{\mathcal{D}}^\alpha_r=\left\{ \mathbf{S}_{-\mathbf{J}}\left(\frac{\p Q^\alpha(\mathbf{x},[\mathbf{u};\gttb])}{\p \mathbf{S}_\mathbf{J}\gtt^r}\right)\right\}\mathbf{S}_{-\mathbf{J}}\,.
\]

\end{theorem}

The proof completely replicates our proof of the continuous version, so we shall omit it.
The identity (\ref{EhrQE}) holds for difference equations, so the set of relations is invariant under locally invertible changes of the arbitrary functions.

\bex\refstepcounter{exple}\label{gaugeex2}
For the interaction of a scalar particle with an electromagnetic field (see Example 1), Christiansen \& Halvorsen (2011) discovered a finite difference approximation that preserves the gauge symmetries as variational symmetries. The mesh $\mathbf{x}(\mathbf{n})$ is uniformly-spaced in each direction, with step lengths
\[
h^\mu=\big(S_\mu-\text{id}\big) \big(x^\mu(\mathbf{n})\big),\qquad \mu=0,\dots,3.
\]
It is useful to introduce the scaled forward difference operators
\[
\ol{D}_\mu=\tl{D}_\mu/h^\mu,\qquad \mu=0,\dots,3,
\]
so that $\ol{D}_\mu$ tends to $D_\mu$ in the limit as $h^\mu$ tends to zero. Note that
\[
\ol{D}_\mu^{\,\bdag}=-\frac{\text{id}-S_\mu^{-1}}{h^\mu}\,;
\]
the adjoint of the (scaled) forward difference operator is the negative of the (scaled) backward difference operator.
Let $\tl{\psi}(\mathbf{n})$ denote the approximation to the wavefunction $\psi$ at $\mathbf{x}(\mathbf{n})$. Christiansen and Halvorsen
used a Yee discretization to approximate $A_\mu$ on the edge that connects the points $\mathbf{x}(\mathbf{n})$ and $S_\mu\mathbf{x}(\mathbf{n})$; we denote this approximation by $\tl{A}_\mu(\mathbf{n})$. As before, indices are raised and lowered using the metric $\eta=\text{diag}\{-1,1,1,1\}$. Up to a sign, the Lagrangian for the scheme is
\begin{equation}\label{emLagd}
L= \frac14 \tl{F}_{\mu\nu}\tl{F}^{\mu\nu} + \left(\tl{\nabla}_{\mu}\tl{\psi}\right)\left(\tl{\nabla}^{\mu}\tl{\psi}\right)^*+m^2\tl{\psi}\tl{\psi}^*
\end{equation}
where, for each $\mu,\nu=0,\dots,3$,
$$\tl{F}_{\mu\nu}=\ol{D}_\mu\tl{A}_{\nu}-\ol{D}_\nu\tl{A}_{\mu},$$
 $$\tl{\nabla}_{\mu}=\frac{1}{h^\mu}\left\{S_\mu-\exp\!\big(\!-\ri e h^\mu \tl{A}_{\mu}\big)\text{id}\right\}=\ol{D}_{\mu}+\frac{1}{h^\mu}\left\{1-\exp\!\big(\!-\ri e h^\mu \tl{A}_{\mu}\big)\right\}\text{id}.$$
Therefore the Euler--Lagrange equations that are obtained by varying $\tl{\psi},\ \tl{\psi}^*$ and $\tl{A}^\sigma$ are
\begin{align*}
 0&= \tl{\mathbf{E}}_{\tl{\psi}} (L) \equiv \tl{\nabla}_\mu^{\bdag}\big(\tl{\nabla}^\mu\tl{\psi}\big)^{\! *}+m^2\tl{\psi}^*,\\
 0&= \mathbf{E}_{\tl{\psi}^*}(L) \equiv \big(\tl{\nabla}_\mu^{\bdag}\big)^{\! *}\tl{\nabla}^\mu\tl{\psi}+m^2\tl{\psi},\\
 0&= \tl{\mathbf{E}}_\sigma (L) \equiv \ \ri e\exp\!\big(\!-\ri e h^\sigma \tl{A}_{\sigma}\big)\tl{\psi}\big(\tl{\nabla}_\sigma \tl{\psi}\big)^{\! *} - \ri e\exp\!\big(\ri e h^\sigma \tl{A}_{\sigma}\big)\tl{\psi}^*\tl{\nabla}_\sigma\tl{\psi}-\eta_{\sigma\alpha}\ol{D}_\beta^{\,\bdag}\tl{F}^{\alpha\beta}.
\end{align*}
Note that $$\tl{\nabla}_\mu^{\bdag}=\ol{D}_{\mu}^{\,\bdag}+\frac{1}{h^\mu}\left\{1-\exp\big(\!-\ri e h^\mu \tl{A}_{\mu}\big)\right\}\text{id}$$ tends to $-\nabla_\mu^*$ as $h^\mu$ approaches zero, so the scheme has the correct limiting behaviour.

The above discretization preserves the gauge symmetries as variational symmetries. Explicitly, these are
$$\tl{\psi}\mapsto \exp(-\ri e\lambda)\tl{\psi},\qquad \tl{\psi}^*\mapsto \exp(\ri e\lambda)\tl{\psi}^*,\qquad \tl{A}^{\sigma}\mapsto \tl{A}^{\sigma}+\eta^{\sigma\alpha}\ol{D}_\alpha\lambda,$$
where now $\lambda$ is an arbitrary real-valued function of $\mathbf{n}$. As in the continuous case, the Lagrangian is invariant, so $\tl{X}L=0$.
The characteristics of the gauge symmetries have components
$$Q^{\tl{\psi}}=-\ri e \tl{\psi}\,\gtt,\qquad Q^{\tl{\psi}^*}\!=\ri e \tl{\psi}^*\gtt,\qquad Q^{\sigma}=\eta^{\sigma\alpha}\ol{D}_\alpha \gtt,$$
where $\gtt$ is an arbitrary real-valued function of $\mathbf{n}$. Consequently the discrete version of Noether's Second Theorem immediately yields the difference relation
\[
-\ri e\tl{\psi}\tl{\mathbf{E}}_{\tl{\psi}} (L)+\ri e\tl{\psi}^*\tl{\mathbf{E}}_{\tl{\psi}^*}(L)+\ol{D}_{\alpha}^{\,\bdag}\left(\eta^{\sigma\alpha} \tl{\mathbf{E}}_\sigma (L)\right)\equiv 0,
\]
which Christensen and Halvorsen obtained with the aid of Noether's first theorem.
\eex

\section{Variational symmetries with difference constraints}

If the functions $\gttb$ are subject to a (complete) set of $S$ linear difference constraints,
\beq\label{constrd}
\tl{\mathscr{D}}_{sr}\gtt^r=0,\qquad s=1,\dots,S,
\eeq
we incorporate these using the same approach as for differential equations. First form the new Lagrangian,
\beq\label{newLcond}
\hat{L}(\mathbf{n},[\mathbf{u};\gttb])=Q^\alpha(\mathbf{n},[\mathbf{u};\gttb])\tl{\mathbf{E}}_{\alpha}(L(\mathbf{n},[\mathbf{u}]))-
\nu^s\tl{\mathscr{D}}_{sr}(\gtt^r),
\eeq
where $\nu^1,\dots,\nu^s$ are Lagrange multipliers. Then take variations with respect to $\gttb$ to obtain
\beq\label{EgQEcond}
\mathbf{S}_{-\mathbf{J}}\left(\frac{\p Q^\alpha(\mathbf{n},[\mathbf{u};\gttb])}{\p \mathbf{S}_{\mathbf{J}}\gtt^r}\,\tl{\mathbf{E}}_{\alpha}(L)\right)\equiv(\tl{\mathscr{D}}_{sr})^{\bdag}(\nu^s),\qquad r=1,\dots,R.
\eeq
Just as in the continuous case, these relations yield conservation laws of the Euler--Lagrange equations (\ref{twelve}), namely
\beq\label{CLd}
\nu^s\tl{\mathscr{D}}_{sr}\gtt^r-\gtt^r(\tl{\mathscr{D}}_{sr})^{\bdag}(\nu^s)=0\qquad \text{on solutions of (\ref{twelve})}.
\eeq
It may also be possible to eliminate the Lagrange multipliers from (\ref{EgQEcond}) to obtain difference relations between the Euler--Lagrange equations.

\bex
The lattice Korteweg--de Vries (KdV) equation is
\beq\label{lkdv}
U_{1,1}-U_{0,0}-c\left(\frac{1}{U_{1,0}}-\frac{1}{U_{0,1}}\right)=0,\qquad c\neq 0,
\eeq
where $U_{i,j}$ denotes the value of the dependent variable $U$ at the point $(n^1+i,n^2+j)$; see Grammaticos \textit{et al.} (1991) for details. This is not an Euler--Lagrange equation as it stands, but it can be turned into one by introducing a potential, $u$, such that
\[
U_{i,j}=u_{i,j-1}-u_{i-1,j}.
\]
Then (\ref{lkdv}) amounts to
\[
\tl{\mathbf{E}}_u(L)\equiv u_{1,0}-u_{0,1}+u_{-1,0}-u_{0,-1}-c\left(\frac{1}{u_{1,-1}-u_{0,0}}-\frac{1}{u_{0,0}-u_{-1,1}}\right)=0,
\]
where
\[
L[u]=u_{0,0}(u_{1,0}-u_{0,1})+c\ln(u_{1,0}-u_{0,1}).
\]
Clearly, the variable $U$ is unaffected by gauge transformations
\[
u_{i,j}\mapsto u_{i,j}+\lambda_{i,j},\qquad\text{where}\quad \lambda_{1,0}=\lambda_{0,1},
\]
so these are symmetries of the Euler--Lagrange equation. Their characteristic is $Q=\gtt_{0,0}$, where $\gtt_{1,0}-\gtt_{0,1}=0$; therefore
\[
\tl{X}(L)=\gtt_{0,0}(u_{1,0}-u_{0,1})=\tl{D}_1\big(\gtt_{\!\,-1,0}\,u_{0,0}\big)+\tl{D}_2\big(\!-\gtt_{0,\!\,-1}\,u_{0,0}\big),
\]
and so the gauge symmetries are variational. Now vary $\gtt$ for the new Lagrangian
\[
\hat{L}[u;\gtt]=Q\tl{\mathbf{E}}_u(L[u])-\nu_{0,0}(\gtt_{1,0}-\gtt_{0,1})
\]
to obtain the relation
\[
\tl{\mathbf{E}}_u(L[u])\equiv \nu_{\!\,-1,0}-\nu_{0,\!\,-1}.
\]
One obvious solution is
\beq\label{nu00}
\nu_{0,0}=u_{0,0}-u_{1,1}+\frac{c}{u_{1,0}-u_{0,1}}\,;
\eeq
the corresponding conservation laws (\ref{CLd}) are
\[
 \tl{D}_1\big(\gtt_{0,0}\,\nu_{\!\,-1,0}\big)+\tl{D}_2\big(\!-\gtt_{0,0}\,\nu_{0,\!\,-1}\big)=0\qquad\text{when}\quad \tl{\mathbf{E}}_u(L)=0.
\]
Note that by setting $\nu_{0,0}=0$ in (\ref{nu00}), we obtain the well-known discrete potential KdV equation,
\beq\label{dpkdv}
(u_{1,1}-u_{0,0})(u_{1,0}-u_{0,1})=c.
\eeq
So the constraint on the gauge symmetry provides a link between the lattice KdV equation and
its related potential form. This observation applies similarly to many integrable
difference equations.
\eex

\section{Conclusion}

We have shown there is a complete correspondence between Noether's Second Theorem for differential and difference equations, and that the same is true where characteristics of variational symmetries depend on partly-constrained functions. Our approach yields, \textit{mutatis mutandis}, the corresponding results for differential-difference equations.

\end{document}